\documentclass[]{aa}  

\usepackage{graphicx, pstricks}
\usepackage{txfonts}
\usepackage{amsmath}
\usepackage{epstopdf}
\usepackage{multicol,lipsum,float}

\newcommand{\dg}{^{\circ}}

\usepackage{natbib}
\bibpunct{(}{)}{;}{a}{}{,}

\begin{document} 

   \title{Near-infrared spatially resolved spectroscopy of (136108) Haumea's multiple system\thanks{Based on observations collected at the European Organisation for Astronomical Research in the Southern Hemisphere, Chile, Program ID: 60.A-9235.}}


   \author{F. Gourgeot
          \inst{1,2}
          \and
          B. Carry\inst{3,4}
          \and
          C. Dumas\inst{1}
          \and
          F. Vachier\inst{3}
          \and
          F. Merlin\inst{5,6}
          \and
          P. Lacerda\inst{7}
          \and
          M. A. Barucci\inst{4}
          \and
          J. Berthier\inst{3}
   }

   \institute{%
     European Southern Observatory, Alonso de C\'{o}rdova 3107, Vitacura, Santiago, Chile.
     \and
     Observat\'orio Nacional, COAA,
     Rua General Jos\'e Cristino 77, 20921-400, Rio de Janeiro, Brazil.
     \email{florian.gourgeot@gmail.com}
     \and
     IMCCE, Observatoire de Paris, PSL Research University, CNRS,
     Sorbonne Universités, UPMC Univ Paris 06, Univ. Lille, France
     \and
     Laboratoire Lagrange, Universit\'e de Nice-Sophia Antipolis, CNRS, Observatoire de la C\^ote d’Azur, France
     \and
     Observatoire de Paris-Meudon / LESIA, Meudon, France.
     \and
     Universit\'e Denis Diderot, Paris VII, France.
     \and
     Max Planck Institute for Solar System Research, Justus-von-Liebig-Weg 3, 37077 G\"{o}ttingen, Germany
   }

 
  \abstract
  {The transneptunian region of the solar system is populated by a
    wide variety of icy bodies showing great diversity in orbital
    behavior, size, surface color, and composition.} 
  {The dwarf planet (136108) Haumea is among the largest
    transneptunian objects and is a very fast 
    rotator ($\sim$3.9h). This dwarf planet displays a highly elongated shape
    and hosts two small moons that are covered with crystalline water ice,
    similar to their central body. A particular region of interest  is the Dark Red Spot  
    (DRS) identified on
    the surface of Haumea from multiband light-curve analysis
    (Lacerda et al. 2008). Haumea is also known to be the
    largest member of the sole TransNeptunian Objects (TNO) family known to date, and an outcome of a
    catastrophic collision that is likely responsible for the unique
    characteristics of Haumea.} 
  {We report here on the analysis of a new set of near-infrared Laser
    Guide Star assisted observations of Haumea obtained with the Integral Field Unit (IFU)
    Spectrograph for INtegral Field Observations in the Near Infrared (SINFONI) at the European Southern Observatory (ESO) Very Large Telescope (VLT)
    Observatory. Combined with previous data published by Dumas et
    al. (2011), and using light-curve measurements in the optical and
    far infrared
    to associate each spectrum
    with its corresponding rotational phase, we were able to carry out a
    rotationally resolved spectroscopic study of
    the surface of Haumea.} 
  { We describe the physical characteristics of the crystalline water ice present on the surface of Haumea for both regions, in and out of the DRS, and analyze the differences obtained for each individual spectrum. The presence of crystalline water ice is confirmed over more than half of the surface of Haumea. Our measurements of the average spectral slope (1.45 $\pm$ 0.82 \% by 100\,nm) confirm the redder characteristic of the spot region. Detailed analysis of the crystalline water-ice absorption bands do not show significant differences between the DRS and the remaining part of the surface. 
    We also present the results of applying Hapke modeling to our data set. The best spectral fit is obtained with a mixture of crystalline water ice (grain sizes smaller than 60 $\mu$m) with a few percent of amorphous carbon. Improvements to the fit are obtained by adding $\sim$10\% of amorphous water ice. 
     Additionally, we used the IFU-reconstructed images to
    measure the relative astrometric position of the largest satellite
    Hi`iaka and determine its orbital elements. An orbital
    solution was computed with our genetic-based algorithm GENOID and 
    our results are in full agreement with recent results.}
  {}


   \maketitle
%

\section{Introduction}

The object (136108) Haumea 
is one of the most remarkable transneptunian objects (TNO). Its
visible light curve indicates a very short ($\sim$3.91 h)
rotation period and an ellipsoidal shape (1000 x 800 x 500 km;
\citealt{Rabinowitz06}), while its gravity is determined to be sufficient for having 
relaxed into hydrostatic equilibrium, making Haumea a dwarf
planet. Dynamically, it is considered a classical TNO, with an
orbital period of 283 years, a perihelion at 35 AU, and an
orbital inclination of 28\degr~\citep{Rabinowitz06}. Haumea is one of the
bluest TNOs \citep{Tegler07} known to date. Its surface is covered by almost pure water ice
\citep{Trujillo07,Merlin07}, though its high density ($\sim$2.6-3.3
g.cm$^{-3}$, \citealt{Rabinowitz06,Lacerda07,Carry12}) indicates a
more rocky interior. It possesses two satellites
\citep{Brown05,Brown06}, the largest of the two is also coated with crystalline
water ice \citep{Barkume06, Dumas11}. 
Such dynamical and physical characteristics, added to the fact that Haumea is the largest 
body of a population of small water-ice TNOs, all sharing the same orbital parameters 
\citep{Schaller08}, point to the idea that Haumea is the remnant body of an ancient (>1 Gyr)
catastrophic collision \citep{Ragozzine07,Snodgrass10,Carry12}. 

High time-resolution observations, combined with multicolor photometry, 
provide evidence for a localized surface feature that is redder and darker than the surrounding material
\citep{Lacerda08,Lacerda09}. The presence of such a feature (perhaps of collisional origin)
makes Haumea the second TNO (after Pluto) displaying strong surface heterogeneities. 
In this study, we combine two sets of observations taken four years apart 
to analyze the spectral behavior of the Dark Red Spot (DRS) and its surrounding areas and the ice properties of the surface material. Finally, we measure the astrometric positions of Hi`iaka directly from our set of hyperspectral images, and determine the Keplerian elements of its orbit using {\em Genoid}, a genetic-based algorithm \citep{Vachier12}. 

\begin{table*}[t]
\caption{\label{hautabdata} Observation logs for Haumea, with its associated central longitude at each epoch, and the solar analog calibrators.}
\begin{tabular}{ccccccc}
\hline
Name&Date \& Time** (UT)&MJD&Longitude***&Exp. Time&Airmass&Seeing\\
of the frames &[yyyy-mm-ss] [hh:mm:ss]&[days]&[$\dg$]&[s]&&["]\\
\hline
H07-0&2007-03-15 06:37:27& 54174.2760&30&300&1.398-1.400&0.81-1.13\\
H07-1&2007-03-15 06:48:40& 54174.2838&47&300&1.398-1.398&0.86-1.05\\
H07-2&2007-03-15 06:54:10& 54174.2876&55&300&1.399-1.398&0.86-1.05\\
H07-3&2007-03-15 07:05:24& 54174.2954&72&300&1.405-1.402&0.84-1.00\\
H07-4&2007-03-15 07:10:53& 54174.2992&81&300&1.409-1.405&0.83-0.98\\
H07-5&2007-03-15 07:16:33& 54174.3032&89&300&1.415-1.410&0.81-0.90\\
\hline
H11-0&2011-04-09 03:52:03 &55660.1612&58&600&1.531-1.498&1.21-1.36\\ 
H11-1&2011-04-09 04:32:30 &55660.1893&120&600&1.425-1.409&0.91-1.25\\
H11-2&2011-04-09 04:53:41 &55660.2040&153&600&1.396-1.387&1.19-1.08\\
H11-3&2011-04-09 05:04:12 &55660.2113&169&600&1.386-1.381&1.03-0.90\\
H11-4&2011-04-09 05:25:26 &55660.2260&201&600&1.379-1.380&0.80-0.80\\
H11-5&2011-04-09 05:35:55 &55660.2333&217&600&1.380-1.385&0.80-1.10\\
H11-6&2011-04-09 07:15:36 &55660.3025&10&600&1.589-1.636&0.89-1.11\\
\hline
HD142093 (G2V) &2007-03-15 07:40:10 &N.A.&N.A.&0.83&1.406-1.408&0.68-0.83\\
HD107087* (G1V) &2011-04-09 01:55:34&N.A.&N.A.&1&1.578-1.639&1.24-1.50\\
HD136776* (G5) &2011-04-09 06:02:45 &N.A.&N.A.&1&1.410-1.431&0.77-0.95\\
BD+192951* (G0)&2011-04-09 06:22:18 &N.A.&N.A.&2&1.397-1.405&0.85-1.26\\
HD146233 (G2V) &2011-04-09 06:43:25 &N.A.&N.A.&0.83&1.083-1.084&0.85-0.86\\
 \hline

\end{tabular} \\
 \\
* Solar analog present in four frames for each 2011 observation. \\
** Middle time of the observation.\\
*** The sub-Earth longitude of Haumea. The origin corresponds to the lowest maximum of the light curve (in the DRS region). \\
\end{table*}

\section{Observations}

  Haumea was observed in H and K bands on 2007 March 15 and 2011 April 09
  UT, using the Laser Guide-Star Facility (LGSF) and the Spectrograph
  for INtegral Field Observations in the Near Infrared (SINFONI)
  instrument installed at the 8-m ``Yepun'' telescope of the
  European Southern Observatory (ESO) Very Large Telescope (VLT). 
  SINFONI is an integral-field spectrometer working in the
  1.0--2.5\,$\mu$m range. It is equipped with an adaptive optics
  (AO) system offering Natural Guide Star (NGS) and Laser Guide Star (LGS)
  channels. The use of this instrument for observing the large
  TNOs Haumea, Eris, and Orcus has been described in earlier papers
  \citep{Merlin07,Barucci08,Dumas07,Dumas11,Carry12}, and more
  information about SINFONI can be found in \citet{Eisenhauer03}
  and \citet{Bonnet04}. The observations for both epochs were carried out using 
  the AO system and its LGS facility. The laser produces an artificial
  visible-light star of R mag\,$\sim$\,13.4 near the line of sight of
  Haumea (V-mag\,=\,17.4), thus providing a gain of four magnitudes for
  characterizing the high orders of the wavefront (in comparison to
  non-laser observations). Haumea was used as the on-axis tip-tilt reference source,
  thus delivering optimal correction by the AO-LGS system.

  \begin{figure}
    \begin{center}
      \includegraphics[width=0.5\textwidth]{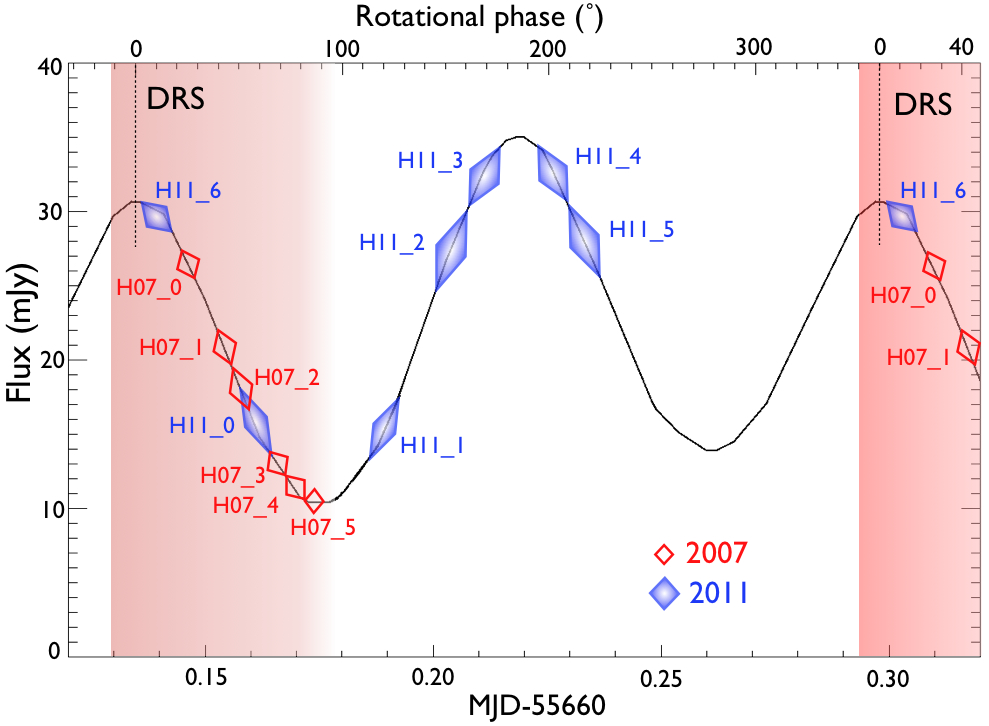}
      \caption{Positions of all observed epochs (six for 2007 and
        seven for 2011, see Table \ref{hautabdata} for notations) onto
        Haumea's light curve. The location of the Dark Red Spot \citep{Lacerda09} is represented 
        by the red-shaded region. The combination of the data obtained on both dates
        returns the 230 degrees of rotational coverage of Haumea. The 2007 observations published 
        in \citep{Dumas11} can be used to characterize the DRS region. 
        The rotational phase (longitude) is also
        represented in degrees along the X-axis. The size of the diamond
        is proportional to the timespan between the start and end of each observation. with
        typical exposure times of 300 s in 2007 and 600 s in 2011. Haumea's light curve is taken 
        from \cite{Lellouch10}. The longitude
        uncertainties are not represented because they are too small
        (less than 0.13\% and 0.07\% over the period covering the 2007 and 2011 data).} 
      \label{haulc}
    \end{center}
  \end{figure}

    \begin{figure*}
    \centering
    \includegraphics[width=0.90\textwidth]{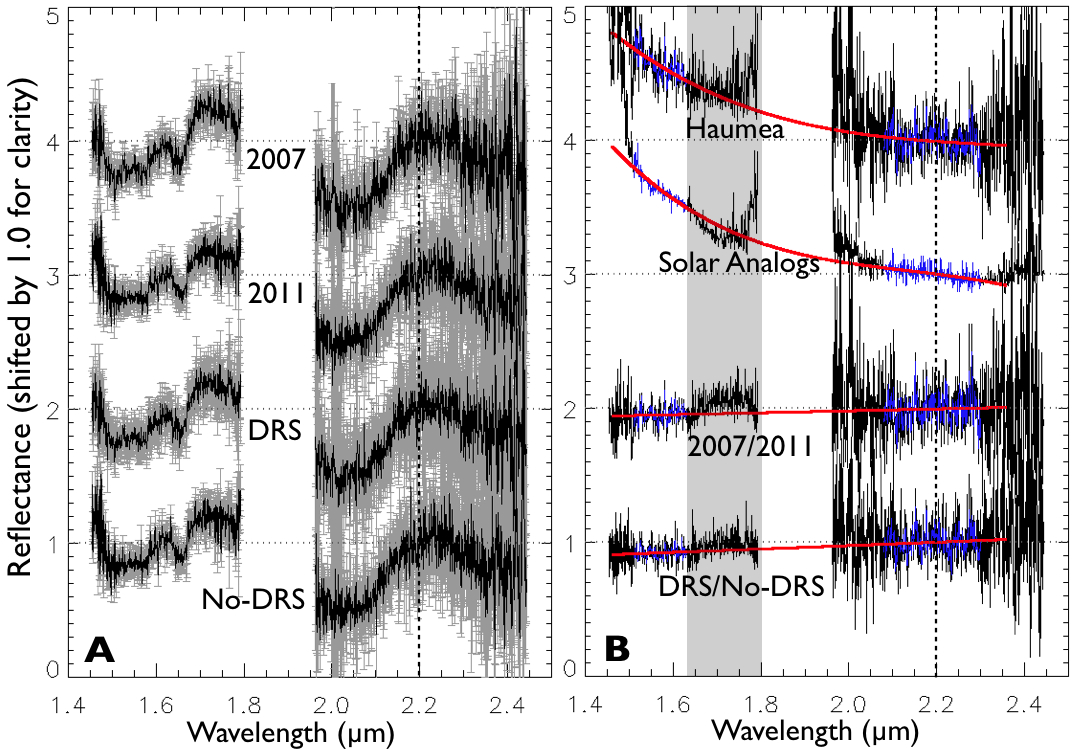}
    \caption{\textbf{A)} Average spectra of Haumea obtained
      for each date (top) and for both regions:      the DRS (including all the 2007
      observations with H11\_0 and H11\_6) and no-DRS (all combined
      H11\_x with x=[1..5]).
      See Table~\ref{hautabdata} and Fig.~\ref{haulc} for notations.
      The standard deviation at 1 $\sigma$
      is also represented for each spectrum. Water-ice  
      features are clearly visible at 1.5 $\mu$m, 1.65 $\mu$m (crystalline), and with the large band 
      at 2.0 $\mu$m, but no significant differences were observed between each regions.
      All spectra were normalized at 2.20 $\mu$m (dashed line).
      \textbf{B)} Spectral ratios between the two data sets of 2007 and 2011: 
      Top: division of the two average spectra of Haumea and that of the two solar
      analogs. We note the bump observed in the 1.63-1.80 $\mu$m range (gray region), 
      which is due to an artifact of the solar analog calibration star
      (see \citealt{Dumas11} for more explanation). 
      Bottom: the ratio of the spectra 2007/2011 and DRS/no-DRS. 
      The blue parts of the spectra correspond to the wavelengths range (1.51-1.63 $\mu$m in H and 2.03-2.30 $\mu$m in
      K) used to fit the
      polynomial regressions (red curves).      
       The slopes are [0.81 $\pm$ 0.60] \%/100 nm and [1.49 $\pm$ 0.60] \%/100 nm  for the 2007/2011 and DRS/no-DRS ratios, respectively, and they confirm the redder profile of the DRS region. 
       }
    \label{hauspectra}
  \end{figure*}
  The atmospheric conditions were extremely good during the 2007
  observations with a seeing varying between 0.81" and 1.13". On
  March 15, between 6h34 UT and 7h24 UT, six exposures of 300 s each
  were obtained on Haumea (total integration time of 0.5 h),
  interspersed with three exposures of 300s to record the sky
  background. The results of these data were already published
  \citep{Dumas11}, but we decided to reduce them again 
  using the same data processing and the latest data reduction tools for both epochs. 
  On 2011 April 09, between 3h57 UT and 5h40 UT, six exposures of 600 s 
  each were obtained on Haumea (total integration time of 1.15 h), interspaced
  by sky background measurements. However, as a result of relatively poor meteorological conditions 
  (degraded sky transparency due to clouds and wind above 15 m/s after 6h00 UT), 
  it was not possible to obtain the full mapping of Haumea during the 2011 observations
  (see Fig. \ref{haulc}). Still, combining the 2007 (sampling almost exactly the 
  DRS longitudinal span) and the 2011 data allowed us to obtain a substantial
  longitudinal coverage of Haumea and explore any variations of its surface properties. 
  The H+K spectral grating (resolving power of $\sim$1500), 
  covering both H and K bands simultaneously, as well as a plate scale of 
  100 mas/pixel (3\arcsec\,$\times$\,3\arcsec~field of view), were used on each observation date.

  Using the same settings (though in NGS), we also observed standard solar-analog stars 
  that are close in time to  observations of Haumea and at similar airmass.
  These stars were used to correct our spectra 
  from the solar response and telluric absorption features 
  (see Table \ref{hautabdata}).

\section{Data reduction}

  The science and calibration data were reduced
  using the ESO pipeline version 2.3.2 \citep{Modigliani07}. We first
  corrected all raw frames from the noise pattern of dark and bright
  horizontal lines introduced when reading the detector.
  We then used the ESO pipeline to produce all master calibration files needed by 
  the data reduction process, such as the bad pixel masks, master darks, flat-field frames, and
  the wavelength and distortion calibration files, which are needed to 
  associate a wavelength value with each pixel, respectively, and to reconstruct the final
  image cubes.

  From each object (science or calibrator) frame, we subtracted the sky frame recorded closest
  in time. The quality of the sky subtraction was further improved by
  enabling the correction of sky residuals in the pipeline, i.e., by
  subtracting the residual median value of the sky from each image slice of the
  spectral-image cube. We extracted all individual spectra one by one and combined them after correction of any remaining bad pixels, 
  which were replaced by the median of all frames at the corresponding wavelength. 
  Airmass correction was applied by dividing the spectra of Haumea with
  their corresponding solar analogs. The final spectra were then normalized to unity 
  at 2.20 $\mu$m. We determined the rotational phase
  of each of our spectra from the accurate knowledge of Haumea's rotation period 
  (3.915341h$\pm$ 0.000005h), based on the closest photometric observations acquired in 
  December 2009 \citep{Lellouch10} and June 2007 \citep{Lacerda09}.
  The 2007 observations were acquired while the DRS region was facing the Earth
 (see Figure \ref{haulc}). An arbitrary origin for the rotational phase of Haumea was chosen 
 to correspond to the lowest of its two light-curve maxima. A detailed analysis of the combined 
  2007 and 2011 observations is shown in the next section.

\section{Spectral analysis}
  \subsection{Identification of the spectral features and comparison}

    Figure \ref{hauspectra}.A shows the average spectra obtained in 2007 and
    2011 in the 1.4--2.4\,$\mu$m range. We also present the spectra
    corresponding to the different regions (DRS and no-DRS).

    These four spectra look broadly similar with no major differences between the DRS and the no-DRS regions, and the main spectral features 
    are that of water ice, with deep absorption bands at 1.5\,$\mu$ and 2\,$\mu$m. 
    The shapes of these bands, and their relative depths between 1.50 and 1.57\,$\mu$m,
    and around 1.65\,$\mu$m, confirm that most of the H$_2$O ice
     is in its crystalline state (amorphous ice lacks the 1.57
    and 1.65\,$\mu$m absorption bands, while 
     the 1.5 and 2.0\,$\mu$m band shapes remain; see \citealt{Grundy98}). This
    predominance of crystalline ice is not surprising, since it is the
    most thermodynamically stable phase of water ice and has 
    been widely reported on the surface of other icy outer solar system objects 
    (e.g.,  satellites of giant planets and transneptunian objects). 

    It is important to note that the spectral difference observed in the 1.63-1.8\,$\mu$m 
    range is due to a feature of the solar analog (HD142093) used for our 2007 observations. 
    It is clearly visible in Fig.~\ref{hauspectra}.A, between 1.63-1.80
    $\mu$m (gray region in Fig. \ref{hauspectra}.B), where we present
    the ratio between both dates for Haumea and the solar analog
    spectra. We decided not
    to remove this artifact in the spectra to conserve all the
    observed properties, but we excluded this wavelength
    region from our analysis of the spectral slope. 
    We could not extract the spectra of the satellites because of their high proximity with Haumea.

  \subsection{Study of the spectral slope}

\begin{figure}
\includegraphics[width=0.5\textwidth]{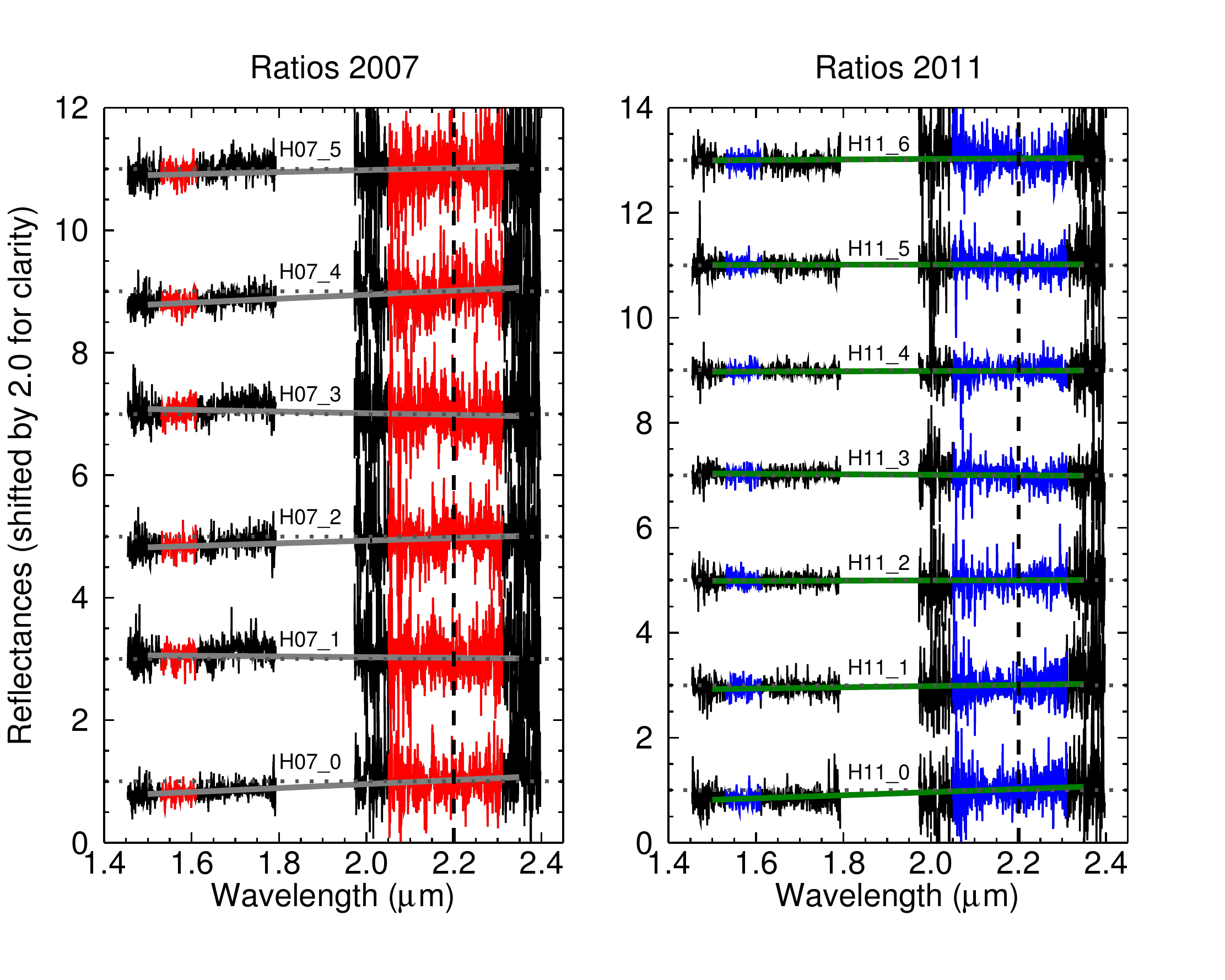}
\caption{
Comparison of the spectra ratios obtained by dividing each spectrum by the average spectrum of 
 the no-DRS region (H11\_x with x=[2,3,4,5]). The slope is calculated considering the
wavelength ranges 1.51-1.63 $\mu$m and 2.03-2.30 $\mu$m (red and blue parts  
for 2007 and 2011, respectively) and a simple linear regression. All spectra have been normalized at
2.20 $\mu$m (dashed line). The variation of the spectral slopes obtained for each spectra as a function of time is represented in Figure \ref{hauslope}.}
\label{haucomparison}
\end{figure}

    As a result of the above analysis, we only considered the spectral 
    ranges between 1.51-1.63\,$\mu$m in H and 2.03-2.30\,$\mu$m in K
    (represented in blue in Fig. \ref{hauspectra}.B) to measure the
    overall spectral slope. We determined a slope of 0.81 ($\pm$ 0.60)
    \%/100 nm for the ratio between the data acquired in 2007 and in 2011, 
    and a slope of 1.49 ($\pm$ 0.60) \%/100 nm for the ratio between 
    the data taken inside and outside the DRS. These values confirm thereby a redder surface
    for the DRS region. This behavior was so far only known for the
    visible part of the spectrum; present measurements are the first
    detection of this redder slope in the near infrared.
    We refined the measurements by computing the ratio
    of each spectrum with the average spectrum of the no-DRS region
    (H11\_x observations, with x = [2,3,4,5]). The spectral slopes are presented 
    in Fig. \ref{haucomparison} and were obtained from a simple linear fit of our data over 
    the same wavelength range as Fig. \ref{hauspectra}.

\begin{figure}
\includegraphics[width=0.5\textwidth]{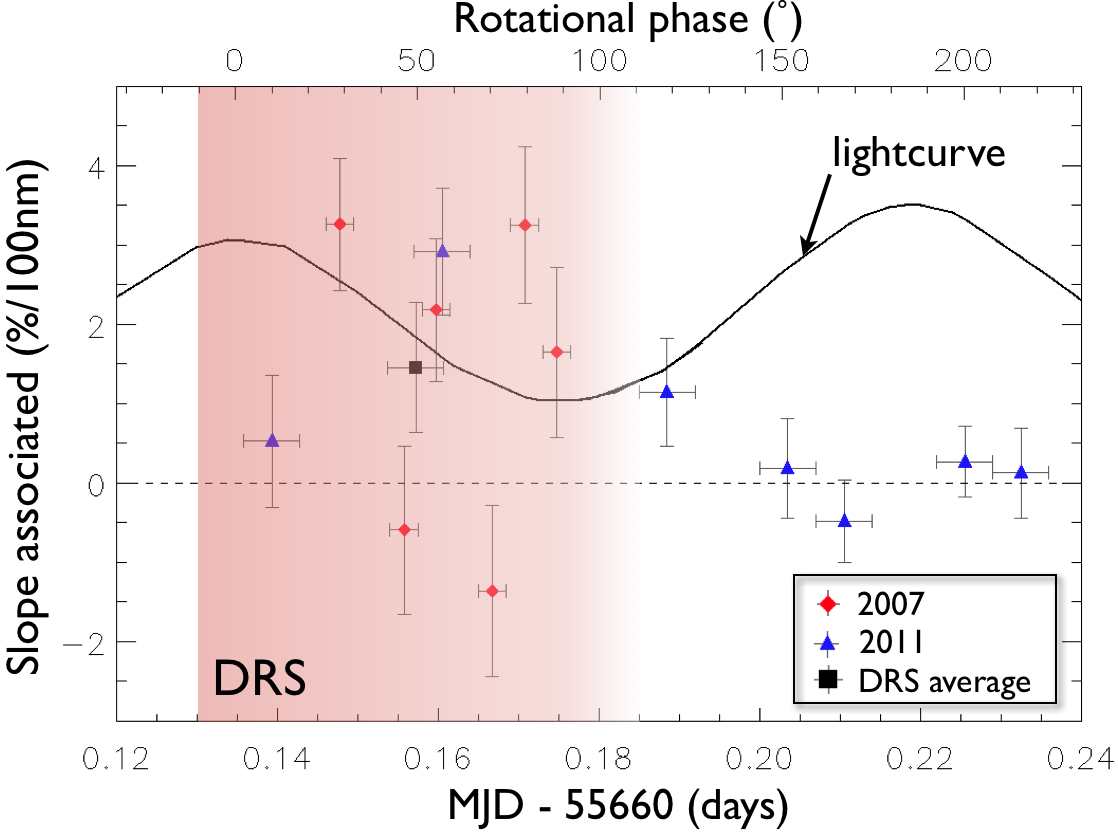}
\caption{Representation of the spectral slopes (\%/100nm)  from Figure \ref{haucomparison} 
  as a function of time (in MJD-55660
  days). The rotational phase (longitude) is also represented on the top axis
  in degrees. The 2007 data have been shifted by an integer
  number of rotational period so that they can be overplotted on the
  2011 data (see Table \ref{hautabdata}). The red circles and blue
  triangles correspond to the 2007 and 2011 data, respectively. A null
  slope corresponding to the no-DRS part (dashed line) and the black square 
  represents the average of the slope values in the DRS equal to [1.46 $\pm$ 0.82] \%/100 nm, 
  highlighting the redder spectral slope for this region of the
  surface of Haumea. 
  The error bars correspond to a standard deviation of 1 $\sigma$. 
  Haumea's light curve (solid line) is also represented for clarity.} 
\label{hauslope}
\end{figure}

    The obtained slope coefficients are presented in Fig.~\ref{hauslope}, which shows the evolution of the spectral slope versus time. 
    The  dispersion of the data points is larger for the DRS region, which is also shown by their larger error bars. 
    Still, the average slope obtained from all DRS measurements point to a redder surface in this region. Similarly, 
    the points with the maximum slope values (3.25 and 3.26 \%/100 nm) are  located at Haumea's longitudes 30$\dg$ and 81$\dg$, 
    respectively, i.e., within the DRS region. 
    Even though better data are needed to improve our measurements, we can confirm from this analysis that the DRS region is spectrally redder than the rest of the surface of Haumea.

  \subsection{Analysis of the 2.0 $\mu$m band depth}

\begin{figure}
\includegraphics[width=0.5\textwidth]{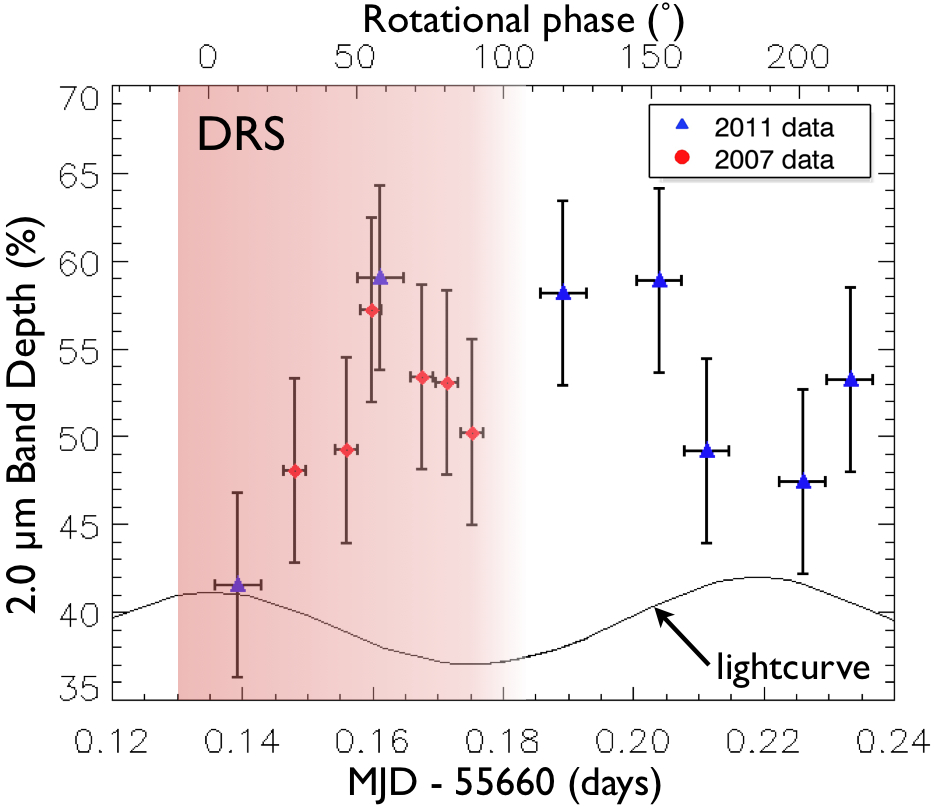}
\caption{Variation of the depth of the 2.0 $\mu$m absorption band (in \%) calculated as a function of time (MJD - 55660 days).  The 2007 data have been shifted by an integer number of rotational period so that they can be overplotted on the 2011 data (see Table \ref{hautabdata}). The red circles and blue triangles correspond to the 2007 and 2011 data, respectively. All spectra were normalized at 2.225 $\mu$m for this study. The error bars are 1-$\sigma$ uncertainty for both data. Haumea's light curve (solid line) is also represented for clarity. Values corresponding to the absorption band depth are in the [40--60\%] range with a smallest water-ice absorption at 10$\dg$ close to the lowest minimum of the light curve with no significant difference between DRS and no-DRS.}
\label{haubd}
\end{figure}

The depth of an absorption feature is computed as
\begin{equation}
D_\lambda[\%] = 100 \times (1 - f_\lambda),
\end{equation}
where $f_\lambda$ is the normalized reflectance at the wavelength $\lambda$. This formula gives the percentile value of the absorption depth of a given band, which informs us about the relative quantity of the absorber and a possible indication of the path length traveled by the light. To compute this parameter we assume the ratio between the median values of the bottom band [2.00-2.05 $\mu$m] and the top band [2.20-2.25 $\mu$m], where all the spectra were normalized (at 2.225 $\mu$m). The results of the band depth are shown in Figure \ref{haubd}. We see that the values of the absorption band depth are in the range 40\% to 60\% with no significant difference between DRS and no-DRS. However, the smallest water-ice absorption (41.5\%) seems to coincide with the smallest longitude (10$\dg$, close to the lowest maximum of the light curve). As a result, the surface material present on the DRS region, even though dominated by crystalline water ice, could correspond to a less hydrated mixture than the remaining  surface of Haumea. 

  \subsection{Analysis of the area of H complex band}

    \begin{figure}
      \includegraphics[width=0.5\textwidth]{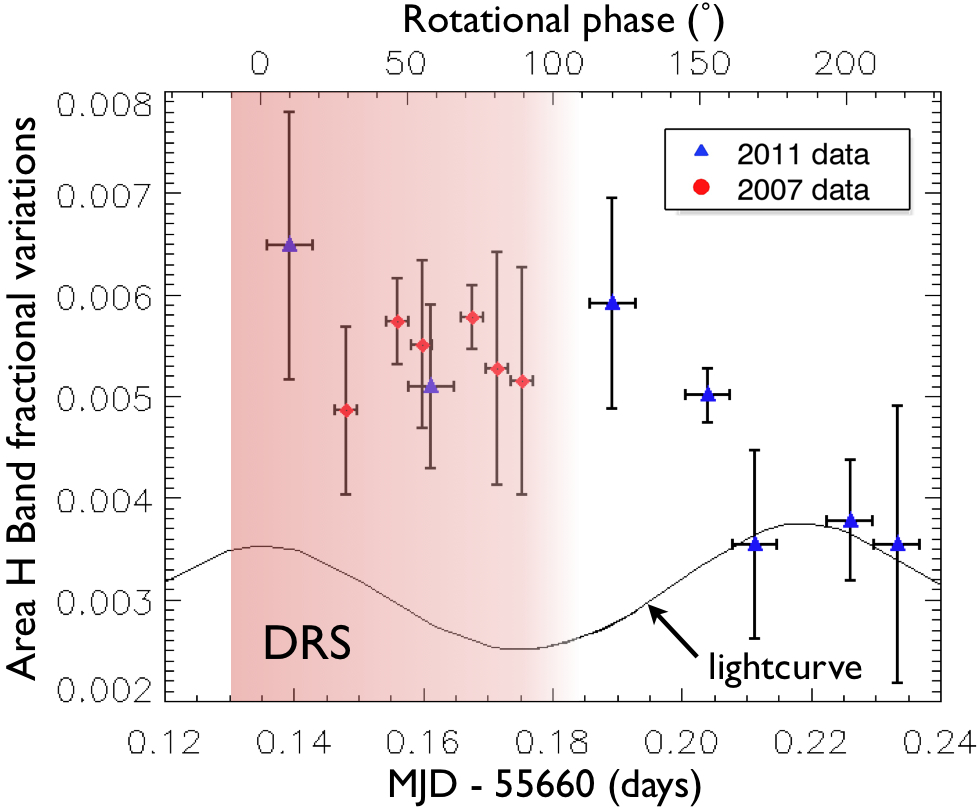}
      \caption{Variation of the integrated area of the [1.50-1.63] $\mu$m H$_2$O ice complex band as a function of time (MJD - 55660 days). The 2007 data have been shifted by an integer number of rotational period so that they can be overplotted on the 2011 data (see Table \ref{hautabdata}). The red circles and blue triangles correspond to the 2007 and 2011 data, respectively. The errors represented are 1 $\sigma$ for both data. Haumea's light curve (solid line) is also represented for clarity. We see that the strongest H-band absorptions are localized in the DRS region.}
      \label{hauarea}
    \end{figure}
    
    However, the most unambiguous characterization of the surface properties of the 
    DRS might come from the spectral analysis of the H-band area. 
    For this, we investigated the behavior of all Haumea spectra around the
    1.50-1.63  $\mu$m absorption band (sadly, we could not study the crystalline
    water-ice band at 1.65 microns because of the artifact introduced by the solar
    analog used in our 2007 data). The integrated
    areas of the 1.5 $\mu$m water-ice absorption complex were computed
    by fitting the continuum level on either side of the absorption band (1.50 to 1.63
    $\mu$m). Then we determined the value corresponding to "one minus the normalized spectrum" over
    the band interval, following a method proposed by
    \citet{Grundy06}. The results are shown in Figure \ref{hauarea}
    where we see the variations of this integrated area with time (and rotational phase). 
    The shallow 1.56 $\mu$m band, still included in our wavelength range of study, is
    also of interest to investigate possible differences between amorphous and
    crystalline ice, as was suggested in the laboratory
    \citep{Mastrapa08}. Figure \ref{hauarea} shows that the strongest absorptions in the H-band region are located within the DRS with a band area almost twice larger (at longitude 10$\dg$) than for the region outside the DRS. Indeed, contrary to the 2.0 $\mu$m band behavior (see previously), the highest
    absorption seems to coincide with the smallest longitude (10$\dg$,
    close to the lowest maximum of the light curve). This result points to the fact that, 
    although the DRS might represent the less hydrated region of the surface, it could still 
    display a higher concentration of crystalline water ice than the rest of the surface. This result itself is
    confirmed by the analysis of our spectral modeling explained in the next section. 

\section{Compositional modeling}

  \subsection{Methods} 
 \begin{table*}
\caption{\label{hautabmix} Model mixtures}
\begin{tabular}{llcclll}
\hline
Models & Compounds & Mixtures (\%) & Grain size ($\mu$m) & Normalization & Slope & ${\chi^2}_{red}$ [1.45-2.45 $\mu$m]\\
\hline
             & Cr. H$_2$O & 95.1 & 40 & & &\\
2007 Model 1 & Cr. H$_2$O & 2.9  & 20 & 0.75 at 1.75 $\mu$m & 0.22 & 1.83\\
             & Am. carbon & 2.0  & 10 & & &\\
\hline
             & Cr. H$_2$O & 87 & 31  & & & \\
2007 Model 2 & Am. H$_2$O & 12.8 & 100 & 0.75 at 1.75 $\mu$m & 0.16 & 1.66\\
             & Am. carbon & 0.2  & 10  & & &\\
\hline
                      & Cr. H$_2$O   & 73 & 9  &&&\\
2007 Previous Model 1 & Am. H$_2$O   & 25 & 10 & 0.60 at 1.65 $\mu$m & N/A & N/A \\
(Dumas et al. 2011)   & Titan Tholin & 2  & 10 & & & \\
\hline
             & Cr. H$_2$O & 86.2 & 55 & & &\\
2011 Model 1 & Cr. H$_2$O & 2.9  & 50 & 0.70 at 1.75 $\mu$m & 0.25 & 1.23\\
             & Am. carbon & 10.9 & 10 & & &\\
\hline
             & Cr. H$_2$O & 82.3 & 57 & & & \\
2011 Model 2  & Cr. H$_2$O & 1.3  & 46 &  0.70 at 1.75 $\mu$m & 0.25 & 1.21\\
        & Am. H$_2$O & 5.9  & 53  & & & \\
            
             & Am. carbon & 10.5 & 10 & & & \\
\hline 
\end{tabular} \\
 \\
"Cr."  and "Am." mean crystalline and amorphous, respectively. 
\end{table*}

    We use the spectral model developed by Hapke (1981, 1993)
    to investigate the chemical properties of the surface of
    Haumea. This approach allows us to model the reflectance spectrum and the albedo
    of a medium, from the physical properties of the different chemical compounds present on the surface.
    The albedo is approximated using Eq. (44) of Hapke
    (1981): 

\begin{equation}
Alb = r_{0}\,(0.5 + \frac{r_{0}}{6}) + \frac{w}{8} \,((1+B_{0})\,P(0)-1)
\label{eq3}
,\end{equation}
where $w$ is the single-scattering albedo and $r_{0}$ the bi-hemispherical reflectance, which is purely single-scattering albedo dependent, i.e.,
\begin{equation}
r{_{0}} = \frac{1-\sqrt{1-w}}{1+\sqrt{1-w}}
\label{eq4}
,\end{equation}
where $w$ depends on the optical constants and the size of the particles and is computed for a multicomponent surface that is assumed to be intimately or geographically mixed; see \cite{Poulet02}. The parameter $B_{0}$ is the ratio of the near-surface contribution to the total particle scattering at zero phase angle and $P$ is the phase function. The method follows the instruction provided by \cite{Merlin10}, assuming an albedo approximation model with a phase angle equal to $0$, and $B_{0}$ close to 0.67 for icy objects \citep{Verbiscer98}. The surface roughness and interference were neglected in this work.
 
\subsection{Results for intimate mixtures}

\begin{figure}
\includegraphics[width=0.48\textwidth]{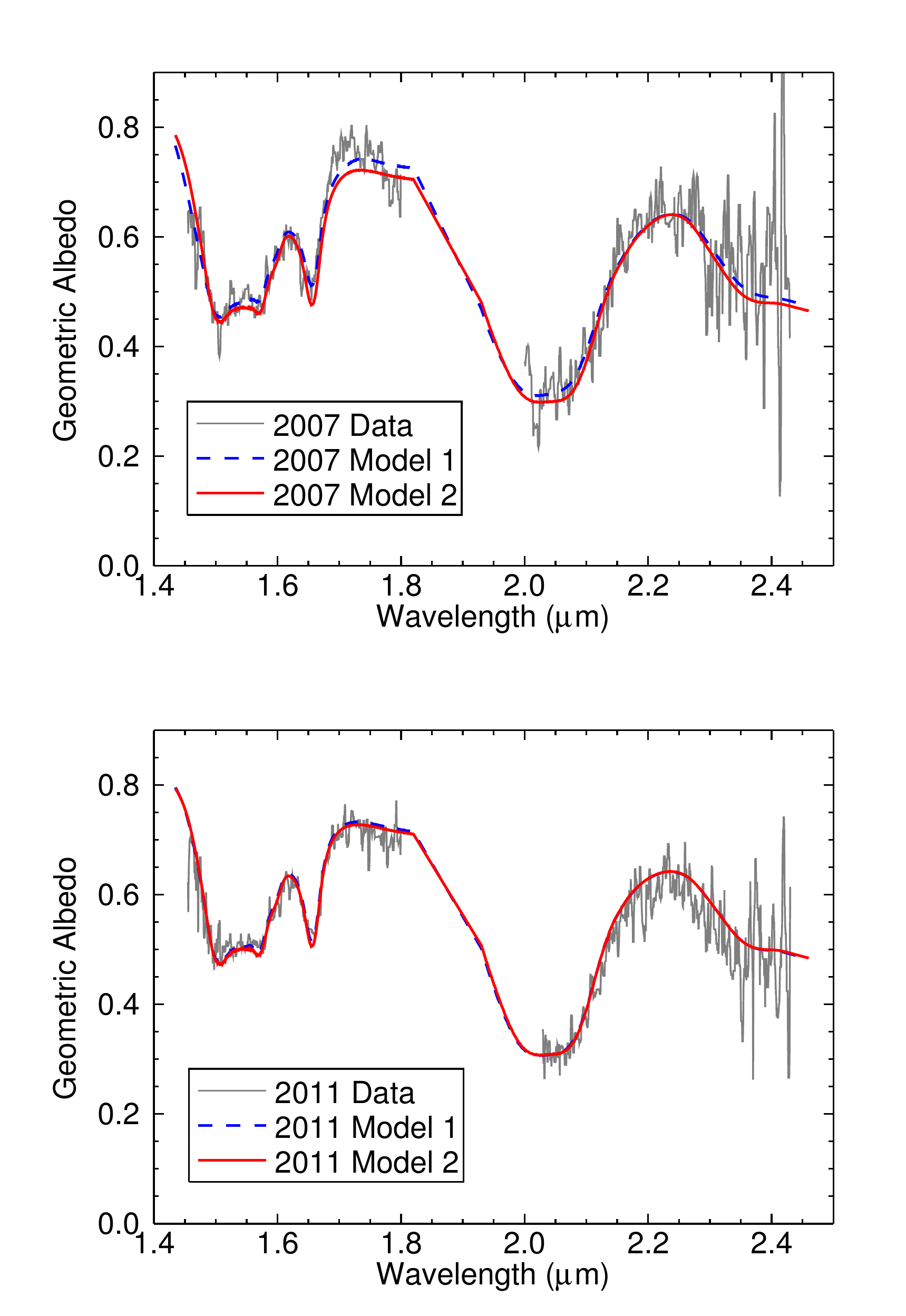}
\caption{Spectra of Haumea (in gray) taken on 2007 (DRS, top figure) and 2011 (no-DRS, bottom figure;  excluding data point H11\_6, which is right on top of the DRS). The results of our spectral modeling is shown with red and blue lines. Crystalline water ice with grain sizes smaller than 60 $\mu$m as the major component, with only a few percent of amorphous carbon, are sufficient to  fit our data  nicely (Model 1, in blue dashed line). However, slight improvements to the fit can be obtained by adding $\sim$10\% of amorphous water ice (Model 2, in solid red line). See Table \ref{hautabmix} for more details on the models.} 
\label{haumix}
\end{figure}

 The free parameters are the asymmetry parameters of the phase function (which is approximated by a single Henyey-Greenstein function), the particle sizes, and the abundances of the various compounds. In order to take  the blue component in our model into account, we add two other parameters, $a$ and $b$, to adjust the blue continuum of the spectrum with a second degree polynomial curve. We note $x$ the wavelength, given by $x$ = ($\lambda$[i]-1.75) $\mu$m, to normalize the spectra at 2.24 $\mu$m with an albedo of 0.67. In each case, the model is adjusted to the spectra using this continuum. The result is shown in Figure \ref{haumix}. The normalization at 1.75 $\mu$m was correlated for each date (geometrical albedo: 0.75 and 0.702 for 2007 and 2011, respectively) taking into account the solar analog artifact described earlier and affecting the 2007 data. A previous model for the 2007 observations \citep{Dumas11} is also represented with a normalization at 0.60 at 1.75 $\mu$m.

 We considered two sets of data: all 2007 observations in a first part (DRS) and all 2011 data (no-DRS, and excluding H11\_6, which corresponds to the DRS region) in the second part. For each date, two models are used: one considering only crystalline water ice, and a second including some amorphous water ice. To constrain the free parameters, we use a best-fit model based on the Levenberg-Marquardt algorithm to minimize the reduced $\chi^{2}$. The minimization is applied outside the telluric absorption bands, which can affect the results. Best models are based on an intimate mixture of crystalline water ice (optical constants at 50K from \citealp{Grundy98}), amorphous carbon (from \citealp{Zubko96}), and amorphous water ice (at low temperature, from \citealp{Mastrapa08}). For the 2007 data, the [1.63-1.80] $\mu$m range was considered but with a lower weight than the rest of the spectra to minimize the contamination of the solar analog artifact. All the results are shown in Table \ref{hautabmix}.\\
Crystalline water ice with grain sizes smaller than 60 $\mu$m as the major component, with a few percent of amorphous carbon, are sufficient to fit the data. Some improvements of the fit can be obtained by adding $\sim$10\% of amorphous water ice. Indeed, we see that, for the 2007 data, the presence of amorphous water ice (with large-size 0.1\,mm grains) allows us to better fit the spectrum of the DRS region, as measured by the small difference of the reduced $\chi^{2}$. For the 2011 data, we obtain a nearly similar reduced $\chi^{2}$ between both models (with and without amorphous ice), although we  get lower values than for 2007 owing to the analog artifact affecting our older data.
 If we focus on Model 2 for both epochs, we confirm our earlier result
 that a higher concentration of crystalline water ice is seen for the
 DRS region (87.0\% for 2007 $vs$ 83.6\% for 2011), considering all
 grain sizes. Finally, we note that the different amounts of amorphous
 carbon needed to fit the 2011 (more than 10\%) and 2007 (less than
 2\%) also points to differences in composition between the DRS and
no-DRS surfaces. It is also important to note that Hershel
   could not detect any ``signature'' associated with the DRS in its
   analysis of Haumea's light curve \citep{Lellouch10}, emphasizing
   the difficulty of quantifying any variation of surface properties at
   the spot's location. 

\section{Orbit of Hi`iaka and Namaka}

\subsection{Astrometric positions}

\begin{figure}[t]
\includegraphics[width=0.48\textwidth]{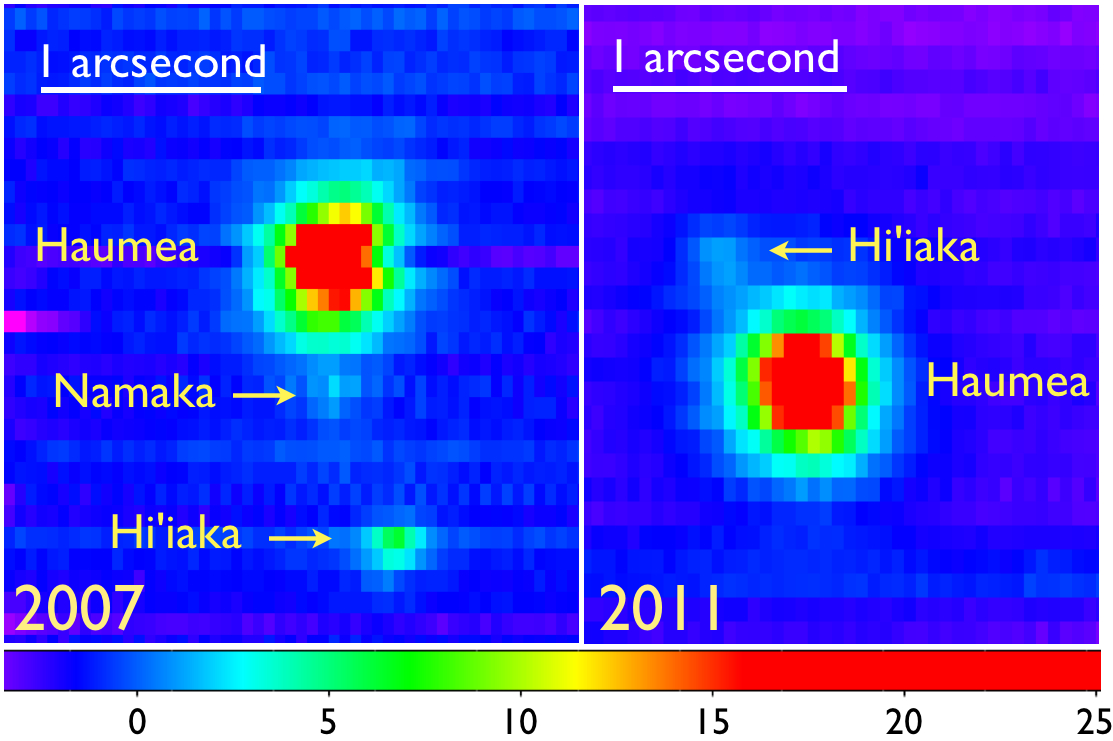}
\caption{Comparison of H+K band of SINFONI broadband images of Haumea obtained 
  with the LGS-AO corrected mode for 2007 (left) and 2011 (right). North is up 
  and east is left. The spatial and intensity scales are similar. The 
  color represents the intensity given in ADU units. The improved contrast 
  and spatial resolution of the AO image make  the detection of the 
  two faint satellites possible: Namaka (the faintest, just below Haumea) and Hi'iaka 
  (the brightest, at the bottom of image) in 2007, only Hi'iaka (faint source
  northeast of Haumea) in 2011.} 
\label{haupos}
\end{figure}

  Taking advantage of the imaging capabilities of SINFONI, we extracted the 
  relative astrometric position of Haumea and its largest satellite Hi`iaka 
  from the hyperspectral cubes of 2172\,x\,64\,x\,64 pixels. We created a broadband 
  image by stacking the cube along the wavelength, avoiding wavelength ranges 
  affected by telluric absorption. The resulting image, shown in Fig.~\ref{haupos}, 
  contains all the slices in the wavelength intervals 1.524--1.764\,$\mu$m and 
  2.077--2.332\,$\mu$m.

\begin{figure}[t]
    \includegraphics[width=0.48\textwidth]{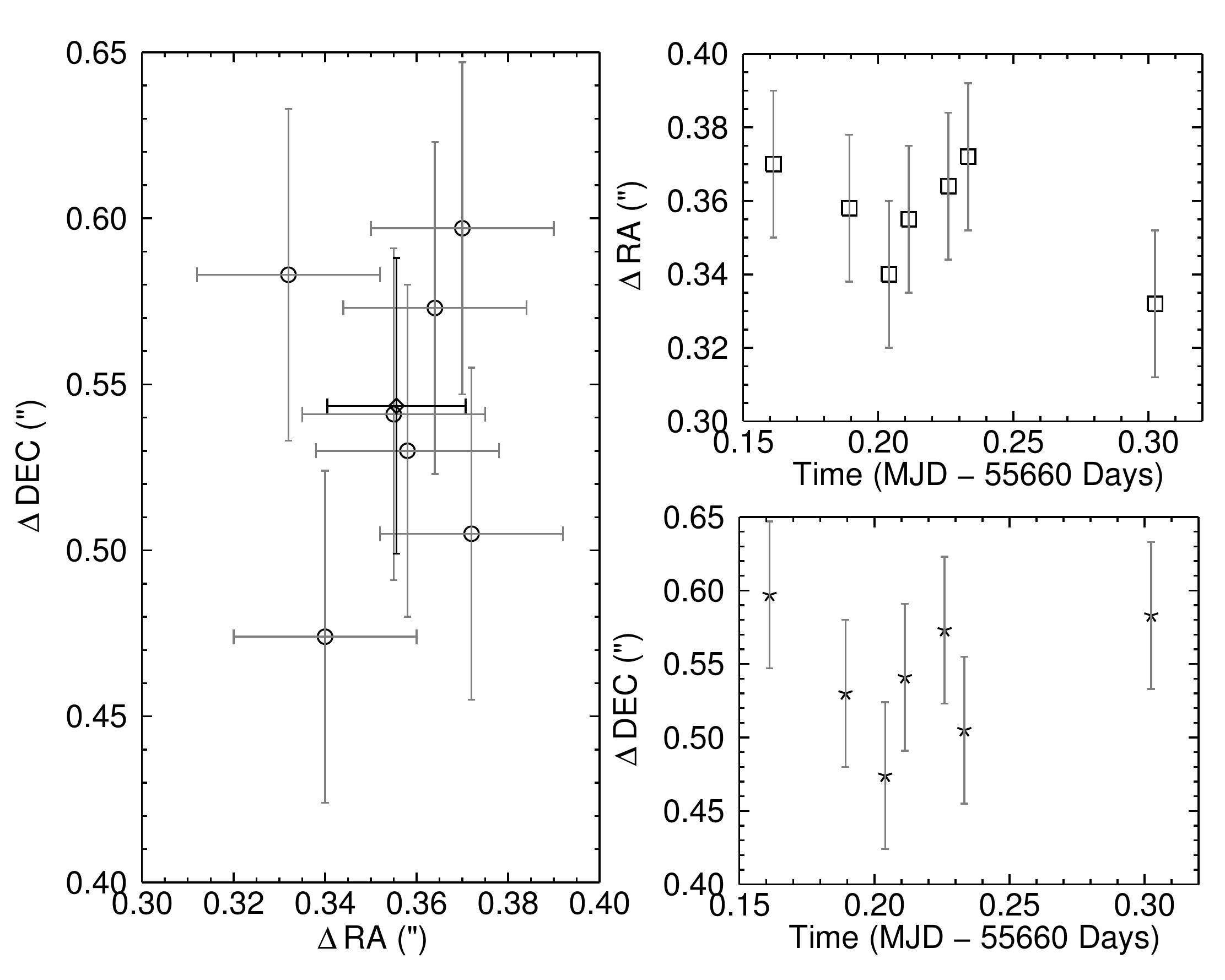}
    \caption{Relative positions of Hi'iaka on 2011 April 09. 
      \textbf{Left:} The seven positions on the plane of the sky
      listed in Table~\ref{haumeas} are shown as open circles, with
      their uncertainty (15x40 mas). The bold diamond represents the
      average position.
      \textbf{Right:} Evolution of the position with time
      (top: $\Delta$RA and bottom: $\Delta$DEC).
      \label{hauposben}
    }
  \end{figure}

  We used Moffat 2-D functions to determine the coordinates of the photocenters 
  of the components, following the procedure described in
  \citet{2011-AA-534-Carry}.
  The spatial accuracy of this measurement is about a quarter of a pixel.
  Owing to the rectangular shape of SINFONI pixels, it is twice as large in the 
  north-south direction (100 mas) than along the east-west direction (50
  mas). The relative positions of Hi`iaka with 
  respect to Haumea photocenter are listed in Table~\ref{haumeas}, and shown 
  in Fig.~\ref{hauposben}. 
  The brightness difference between Hi`iaka and Haumea, which is normalized to
  the average brightness of Haumea, that is, removing the effect of
 Haumea's high-amplitude light curve, is also reported as a magnitude
  difference ($\Delta$Mag) for each epoch.
  With the exception of the last epoch, the magnitude difference
  ranges from 2.6 to 3.1, with an average of 2.76\,$\pm$\,0.18, in agreement
  with the value of 3.0 reported in \citet{Lacerda09}. 
  The estimation of the flux of the components is rather crude, and is 
  reported here only for information.
  Namaka is not detected in 2011 image (see Fig.~\ref{haupos}), 
  although its detection is within SINFONI capabilities \cite[see Fig.~\ref{haupos}
  and][]{Dumas11}. It is likely that Namaka was angularly too close to Haumea at 
  the time of the observations, as suggested by its predicted position at
  (0\arcsec.008, -0\arcsec.110), based on the orbit by
  \citet{Ragozzine09} and our own computations below, 
  thus less than 2 pixels from the center of Haumea. 

  \begin{table}[t]
    \caption{\label{haumeas} Relative positions of Hi'iaka on 2011 April 09. 
        The uncertainty is 15 mas in right ascension, and 40 mas in declination.}
    \begin{center}
      \begin{tabular}{ccccc}
        \hline
        UT &    $\Delta$RA (\arcsec)   & $\Delta$DEC (\arcsec)  &  Flux ratio &  $\Delta$Mag \\
        \hline
        03:52:03 & 0.370 & 0.597  &  0.0956 & 2.9 \\
        04:32:30 & 0.358 & 0.530  &  0.0865 & 3.1 \\
        04:53:41 & 0.340 & 0.474  &  0.0699 & 2.6 \\
        05:04:12 & 0.355 & 0.541  &  0.0618 & 2.6 \\
        05:25:26 & 0.364 & 0.573  &  0.0590 & 2.6 \\
        05:35:55 & 0.372 & 0.505  &  0.0619 & 2.7 \\
        07:15:36 & 0.332 & 0.583  &  0.0177 & 4.1 \\
        \hline
      \end{tabular}
    \end{center}
  \end{table}

\subsection{Orbital solutions}

 \begin{figure*}
      \centering
      \includegraphics[width=0.75\textwidth]{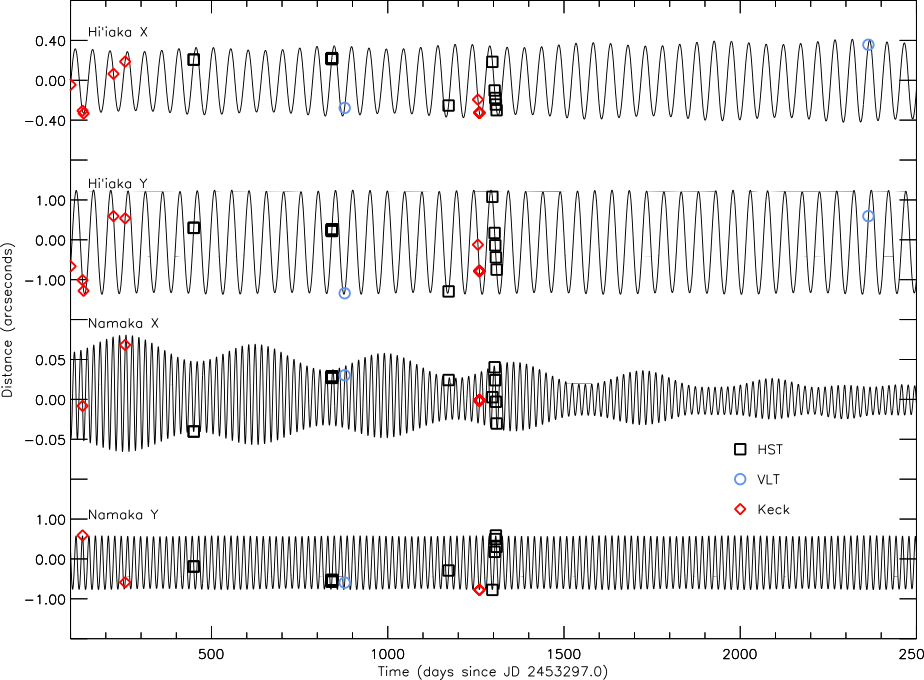}
      \caption{Observed positions and model positions of Hi`iaka and
        Namaka. Figure similar to Fig. 2 of \citet{Ragozzine09}, with similar
        scale to ease comparison.
        From top to bottom, the curves represent the model
        on-the-sky position of Hi`iaka in the x-direction (i.e., the
        negative offset in right ascension), Hi’iaka in the
        y-direction (i.e., the offset in declination), Namaka in the
        x-direction, and Namaka in the y-direction, all in arcseconds.
        Symbols represent astrometric observations. We note the much
        larger time coverage provided by the use of both Keck and
        VLT/SINFONI data set.
        \label{fig: orbit_v_t}
      }
    \end{figure*}
    
  \indent \citet{Ragozzine09} built a dynamic solution for Haumea's system
  on the base of about 30 observations spanning a period of 1260 days from
  January 2005 to May 2008, acquired with the Hubble Space Telescope (HST).
  These authors used a dynamic three-body model in which the gravitational interaction 
  between the satellites causes non-Keplerian perturbations on Namaka's orbit 
  on timescales much longer than a month. However, they found that the 
  perturbation of Namaka on Hi`iaka's orbit was negligible, and that its orbit is 
  described well  by a Keplerian motion. Furthermore, they found no evidence 
  for an effect of Haumea gravitational quadrupole ($J_2$). Using the 
  astrometric positions of Hi`iaka and Namaka measured on SINFONI datacubes 
  in 2007, \citet{Dumas11} also studied the orbits of both satellites and 
  discussed heating that could be generated by tidal effects. \\
  \begin{table*}[!t]
    \begin{center}
      \caption{\label{tab:hiiaka}
        Dynamical parameters of Hi`iaka and Namaka 
        reported in EQJ2000 and derived mass of Haumea.
      } 
      \begin{tabular}{lcccccl}
        \hline
        \hline
        & \multicolumn{2}{c}{Hi`iaka} &\multicolumn{2}{c}{Namaka}\\
        Parameter & Value & 1-$\sigma$ & Value & 1-$\sigma$ & Units &Comment\\
        \hline
        $P$           &      49.031527 &   0.008980 &     18.323535 &   0.003016 & day  & Orbital period\\
        $a$            &  49502.940    & 741.272    &  25147.829    & 634.162    & km   & Semi-major axis \\
        $e$            &      0.05260  &   0.00599  &      0.15543  &   0.03127  & --   & Eccentricity \\
        $i$            &    259.48     &   0.71     &     88.83     &   1.13  & deg. & Inclination \\
        $\Omega$       &    192.99     &   0.21     &    206.73     &   0.67 & deg. & Longitude of ascending node \\
        $\omega$       &    276.14     &  10.40     &    143.44     &  11.07        & deg. & Argument of the pericenter \\
        $t_{p}$ (JD)   & 2452190.3944  &    1.383   & 2452167.5299  &   0.763 & day  & Time of pericenter\\
        Mass           & 3.999         & 0.179      &  3.754        & 0.284   & $\times 10^{21}$ kg & System mass \\
        \hline
      \end{tabular}
    \end{center}
  \end{table*}
  \indent We took advantage of another astrometric position, taken 431 days 
  (about nine revolutions) after the latest position reported by \citet{Ragozzine09} 
  to determine the orbital elements of Hi`iaka. For that, we used 
  the genetic-based algorithm \textit{Genoid} \citep{Vachier12}, which
  relies on a metaheuristic approach to find the best-fit set of
  orbital elements in a two-body problem.
  This approach can be used to search for Keplerian orbits, as 
  well as to explore a more complex problem, including the gravitational 
  field of the central body up to the fourth order. To define the set of orbital 
  elements that best fit the data, \textit{Genoid} minimizes a fitness function, 
  $f_p$, defined as a $\chi^2$ minimization function (eq. \ref{eq:chi2})
  \begin{equation}
    \label{eq:chi2}
    f_p = \chi^2 = \sum_{i=1}^{n} \left[ \left(\frac{x^{o}_i - x^{c}_i}{x^{e}_i} \right)^2 + 
      \left(\frac{y^{o}_i - y^{c}_i}{y^{e}_i} \right)^2 \right]
  ,\end{equation}

  where $n$ is the number of observations, and $x_i$ and $y_i$
    are the relative position between 
  Hi`iaka and Haumea along the right ascension and declination, respectively.
  The exponents $o$ and $c$ stand for observed and computed positions, and 
  exponent $e$ stands for measured error. For convenience, in the following 
  text, we express the fitness function as the quadratic mean of the residuals.
  The main advantage of this fitness function is to provide a link between 
  the quality of the fitted orbit and the uncertainties of the astrometric 
  positions provided by the ephemeris. A definition and discussion about estimating 
  uncertainties in astronomy can be found, for example, in \cite{Andrae10},
  \cite{Andrae10b}, and references therein.\\
  \indent We used a total of 35 astrometric positions, spread over 2264 days
  or 46.2 orbital periods, to determine the orbit of Hi`iaka. The
  orbit of Namaka is based on 30 observations spread over 1175 days
  or 64.1 orbital periods.
  We added to the set of HST observations, the positions measured on SINFONI spectral cubes in 2007
  and 2011 (see above) and the
  set of observations obtained with NIRC2 on the
  W. K. Keck II telescope (pixel scale of 9.963 mas) between
  2005 and 2008 reported by 
  \citet{Brown05}, \citet{Brown06}, and \citet{Ragozzine09}.\\
  \indent Albeit the intrinsic precision of Keck images is deemed cruder than
  that of HST, which benefits from highly stable PSF, and the absolute
  link between HST and Keck astrometry is unknown (pixel scale
  and field orientation), we favor orbital solutions based on a longer
  time baseline that are more sensitive to departures from a pure
  Keplerian motion. Because of these points, however, the goodness of
  fit is expected to be lower.
  The addition of Keck data here corresponds to an increase of 350
  days of time coverage. Compared to the case using only HST data,
  as carried out by \citet{Ragozzine09}, this represents 44\% of the temporal
  baseline.
  This baseline is longer with the addition of the SINFONI
  measurements in 2011, the leverage provided by Keck data is of 20\%
  (see Fig.~\ref{fig: orbit_v_t}).\\

\subsection{Results}

  \indent We run two separate orbital fits: one assuming the central object
  (Haumea) to be a point-like mass and the satellites to be massless 
  test particles (i.e., \textit{Genoid-Kepler}), and one considering the zonal 
  harmonic coefficients $J_{2}$ and $J_{4}$ of the central object, its 
  size, and the coordinates of its pole of sidereal rotation (i.e.,
  \textit{Genoid-ANIS}). 

    \indent Our best-fit solution is obtained using a pure Keplerian motion with a
    fitness function $f_p$\,=\,11.3\,mas for Hi`iaka and 
    $f_p$\,=\,17.3\,mas for Namaka. This 
    corresponds to reduced $\chi^2$ of
    1.61 and 2.5, respectively.
    This level of accuracy is typical of the pixel size of the Keck NIRC2
    camera of $\approx$10\,mas, which is 5 and 10 times smaller than
    the pixel size of HST WFPC2 ($\approx$50\,mas) and ACS
    ($\approx$100\,mas) cameras, respectively. 
    These reduced $\chi^2$ are to be compared with the reduced $\chi^2$
    of 1.1 reported by \citet{Ragozzine09}, based on a fit of both HST
    and Keck data.\\
    \indent Our calculations show clearly that there is no signature for
    non-Keplerian motion that would be 
    caused by Haumea gravitational quadrupole, confirming
    \citet{Ragozzine09} results.
    Although the orbits determined here are restrained to the two-body
    assumption, compared with the three-body integration by 
    \citet{Ragozzine09}, they still allow us to predict the position of
    both satellites to a high accuracy over the period 2005-2020.
    Our Virtual Observatory web service
    Miriade\footnote{\href{http://vo.imcce.fr/webservices/miriade/?ephemcc}{http://vo.imcce.fr/webservices/miriade/?ephemcc}}
    \citep{2009-ESPC-Berthier}
    allows anyone to 
    compute the ephemeris of Hi`iaka and Namaka for any arbitrary
    epoch and observer location, and can be used to plan future
    observations of the system.

\begin{figure}
      \includegraphics[width=0.48\textwidth]{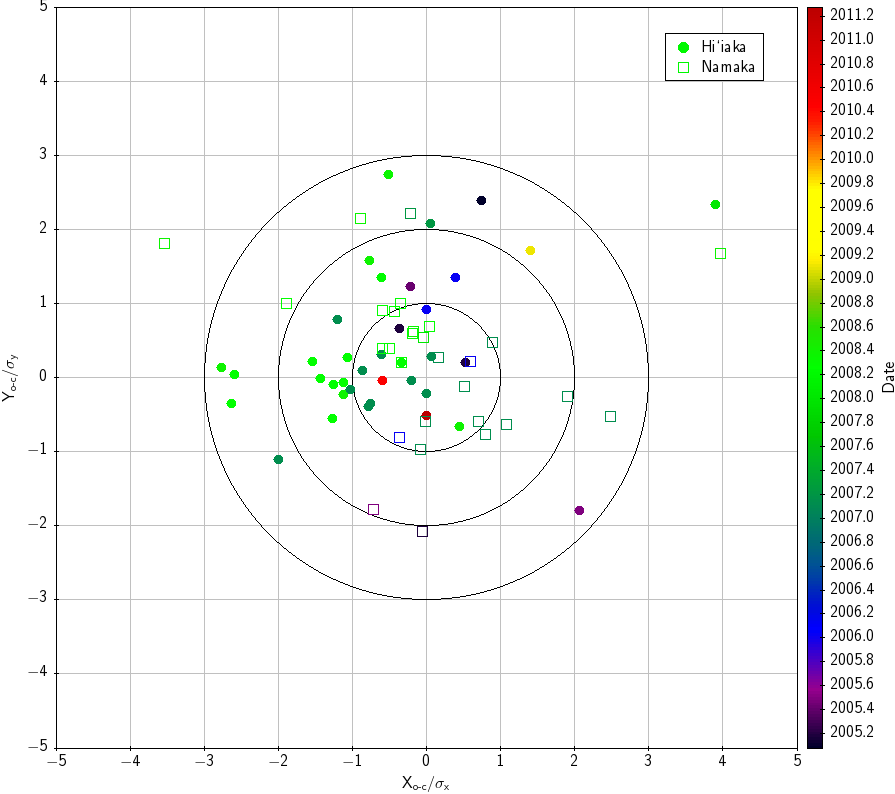}  
      \caption{Residuals (observed minus computed satellite -- primary positions) 
        normalized by the positional uncertainty $\sigma$ for both
        Hi`iaka and Namaka (filled and open symbols, respectively). The three circles represent 
        the 1-, 2-, and 3--$\sigma$ contours. The color enlightens the different epochs
        of observation. The redder dot corresponds to the observation by SINFONI in 2011.
        \label{fig:hauhianomc}
      }
    \end{figure}

 Figure \ref{fig:hauhianomc} shows the residual positions (between observed and computed) obtained normalized by the positional uncertainty. We list in Table \ref{tab:hiiaka} the orbital parameters 
    of Hi`iaka and Namaka obtained with \textit{Genoid-Kepler}. 
    They correspond to the same geometry as the orbits presented by
    \citet{Ragozzine09}, albeit Namaka eccentricity is found to be lower
    (0.15 here vs 0.25). We note the existence of a numerical solution
    with similar residuals ($f_p$\,=10.7\,mas) in which the orbital
    motion of Hi`iaka is reversed. We do not consider this solution here.
    
\section{Conclusion}

  We presented a rotationally resolved spectroscopic study of the surface of the transneptunian object 
  Haumea derived from the 2007 and 2011 near-infrared (1.4--2.5 $\mu$m) observations obtained with
  the integral-field spectrograph SINFONI at the ESO VLT. Crystalline 
  water ice is confirmed to be the major compound present over the entire surface of Haumea.
  We detected a steeper spectral slope in the near-infrared
    associated with the so-called DRS region over
    which the visible spectrum is redder, as reported by
    \citet{Lacerda09}. We find that the reddening of the spectrum 
    extends up to 2 microns.  
  
  A detailed analysis of the depth of the 2.0\,$\mu$m absorption band
  shows that the weakest water-ice absorption is measured for 10$\degr$ of 
  rotation phase, i.e., near the center of the DRS region. 
  Similarly, the analysis of the water absorption band around
  1.6\,$\mu$m reveals that the concentration
  of crystalline water ice is highest at the DRS location.
  Hapke modeling of our spectra also shows that amorphous carbon appears to
  be depleted in the DRS region in comparison to the rest of the surface.
  Although no major compositional differences have been
    found between the DRS region and the remaining
    surface, all the evidence hints at a slightly different composition
    for this region.

  New astrometric positions of Hi`iaka were reported and an orbital
  solution was computed with the genetic-based algorithm
  \textit{Genoid}. The results are in agreement with the orbits
  originally presented by \citet{Ragozzine09} and ephemeris generation
  is proposed to the community.


\begin{acknowledgements}
  We thank Darin Ragozzine for providing the expected positions of
  both satellites from his modeling at the time of our observations.
\end{acknowledgements}



\bibliographystyle{aa}

\label{lastpage}

\end{document}